\documentclass[twocolumn,showpacs,aps,prl]{revtex4}
\usepackage{epsf,psfig}

\begin{document}

\title{
Mid-rapidity $\Lambda $ and $\overline{\Lambda }$ Production in Au+Au
Collisions at $\sqrt{s_{_{NN}}}=130$ GeV}

\author{
\begin{flushleft}
C.~Adler$^{11}$, Z.~Ahammed$^{23}$, C.~Allgower$^{12}$, J.~Amonett$^{14}$,
B.D.~Anderson$^{14}$, M.~Anderson$^5$, G.S.~Averichev$^{9}$, 
J.~Balewski$^{12}$, O.~Barannikova$^{9,23}$, L.S.~Barnby$^{14}$
% ...
\end{flushleft}
\begin{center}(STAR Collaboration)\end{center}
}

% ...
\affiliation{$^5$University of California, Davis, California 95616}
% ...
\affiliation{$^{9}$Laboratory for High Energy (JINR), Dubna, Russia}
% ...
\affiliation{$^{11}$University of Frankfurt, Frankfurt, Germany}
\affiliation{$^{12}$Indiana University, Bloomington, Indiana 47408}
% ...
\affiliation{$^{14}$Kent State University, Kent, Ohio 44242}
% ...
\affiliation{$^{23}$Purdue University, West Lafayette, Indiana 47907}
% ...

\date{\today}

\begin{abstract}
We report the first measurement of strange ($\Lambda$) and
anti-strange ($\overline{\Lambda}$) baryon production from 
$\sqrt{s_{_{NN}}}=130$ GeV Au+Au collisions at the
Relativistic Heavy Ion Collider (RHIC). Rapidity density and
transverse mass distributions at mid-rapidity are presented as a function of centrality.
The yield of $\Lambda$ and $\overline{\Lambda}$ hyperons is found to be 
approximately proportional to the number
of negative hadrons. The production of $\overline{\Lambda}$ hyperons relative to negative hadrons 
increases very rapidly with transverse momentum. The magnitude of the increase cannot be described by
existing hadronic string fragmentation models.
\end{abstract}

\maketitle

Ultra-relativistic nucleus-nucleus collisions provide a unique means
to create nuclear matter of high energy density (temperature) and/or
baryon density over an extended volume~\cite{Blaizot}.  The first
results from the Relativistic Heavy Ion Collider (RHIC) have shown
that the large charged particle multiplicity, measured in Au+Au
collisions at $\sqrt{s_{_{NN}}}=130$~GeV, corresponds to an energy
density significantly higher than that previously achieved in heavy
ion collisions~\cite{mult,phenix,starh}. In addition, the anti-proton
to proton ratio at mid-rapidity has been measured to be in the range
$0.6$--$0.7$~\cite{ratio,phobos,brahms}, which is indicative of particle
production from a low net baryon density regime. Thus the global 
characteristics of nucleus-nucleus collisions at RHIC are the formation
of a high energy density and low net baryon density region at mid-rapidity.

The yield of baryons and anti-baryons produced in relativistic nuclear collisions 
is very important because it is sensitive to two fundamental, not yet fully understood 
aspects of hadron production dynamics: baryon/anti-baryon pair production and the transport of baryon 
number from beam rapidity to mid-rapidity.
The nature of baryon/anti-baryon production itself is the subject of much interest, 
with theoretical conjecture addressing a range of possible mechanisms from string 
fragmentation~\cite{jetset}, to exotic mechanisms involving Quantum 
Chromo Dynamics (QCD) domain walls~\cite{Ellis}. 
The physical nature of the entity which carries baryon number, and the means by which 
baryon number is transported 
over a large rapidity gap into the mid-rapidity region are also subjects of considerable 
experimental and theoretical interest~\cite{Ledoux,Baryon,Khar}.

Strange baryon/anti-baryon production is particularly interesting due to the increased 
sensitivity to the availability of strange/anti-strange quarks, which is expected to be 
suppressed relative to light quarks in hadronic matter due to the strange quark mass.
Strangeness production has long been predicted to be a signature of Quark Gluon Plasma (QGP)
formation~\cite{Rafelski}.  The strangeness production in previous
generations of heavy ion experiments has been observed to be
significantly increased compared to those from p+p, p+A and light ion
collisions~\cite{E802,Ahm,WA97,NA49}, although questions remain about the
exact strangeness production mechanism. In particular, the relative 
importance of strange baryon
production from hadronic rescattering differs between
calculations~\cite{Koch,Rene}, depending on both the evolution of the
system and the scattering cross sections assumed.  Exotic dynamical
mechanisms that have been proposed for strange baryon
production include, for example, Color String Ropes~\cite{Sorge},
String Fusion~\cite{Amelin} and Multi-mesonic Reactions~\cite{Rapp}.
All require a high local energy density and therefore suggest that
strangeness production occurs early in the collision.

In this letter we report on mid-rapidity ($|y|$$<$$0.5$) Lambda
($\Lambda $) and anti-Lambda ($\overline{\Lambda}$) production in
Au+Au collisions at $\sqrt{s_{_{NN}}}=130$ GeV. The data were taken
with the STAR (Solenoidal Tracker At RHIC) detector. The main components of the detector 
system for this analysis have been described in detail elsewhere~\cite{Flow}. They included a
large volume Time-Projection Chamber (TPC)~\cite{TPC}, a pair of
Zero-Degree Calorimeters (ZDC) located at +/- 18 meters from the center of the TPC, 
and a Central Trigger Barrel (CTB) constructed of scintillator paddles surrounding the TPC. 

The TPC was used to provide tracking information and particle identification by measuring 
the specific ionization loss (dE/dx) for charged particles. The ZDCs were used in 
coincidence to define a minimum bias trigger by selecting interactions from the intersection
region of the colliding beams, and to measure the forward 
going energy of near-beam rapidity fragments, consisting primarily of spectator neutrons. 
The CTB was used to define a central interaction trigger based on the energy 
deposition from charged particles entering its acceptance.

Data from both the minimum bias trigger and central trigger have been
used for this analysis. Similar to a previous analysis~\cite{starh},
the collision centrality was defined off-line using the total charged particle
multiplicity within a pseudo-rapidity window of $\left| \eta\right|
<0.5$. The charged particle multiplicity distribution was divided into
five centrality bins, corresponding to approximately the most central $5\%$,
$5$--$10\%$, $10$--$20\%$, $20$--$35\%$, $35$--$75\%$ of the total
hadronic inelastic cross section of Au+Au collisions.

The $\Lambda $ and $\overline{\Lambda }$ particles were reconstructed
from their weak decay topology, $\Lambda \rightarrow p\pi ^{-}$ and
$\overline{\Lambda }\rightarrow \overline{p}\pi ^{+}$, using charged
tracks measured in the TPC. These tracks were projected back from
the TPC to find their Distance of Closest Approach (DCA) to the primary vertex. 
Since the tracks of particles from $\Lambda $ and $\overline{\Lambda}$ decays 
should not originate from the primary vertex, only proton (anti-proton) candidate 
tracks missing the primary vertex by 0.90 cm and $\pi^{-}$ ($\pi^{+}$) candidate 
tracks missing the primary vertex by 2.85 cm were selected.
Particle assignments for $p$
($\overline{p}$) and $\pi ^{-}$ ($\pi ^{+}$) candidates were based on
charge sign and the mean energy loss, $<$$dE/dx$$>$, measured for each
track.  Candidate tracks were then paired to form neutral decay
vertices, which were required to be at least 5 cm from the primary vertex. The
momentum vector of the reconstructed parent particle was required to
originate from the primary interaction: the distance of closest approach of the
parent trajectory to the primary vertex was required to be less than
0.5 cm. The effect of these cuts on the reconstruction efficiency was 
determined using the embedding techniques described below.

\begin{figure}
\centering\mbox{
\psfig{figure=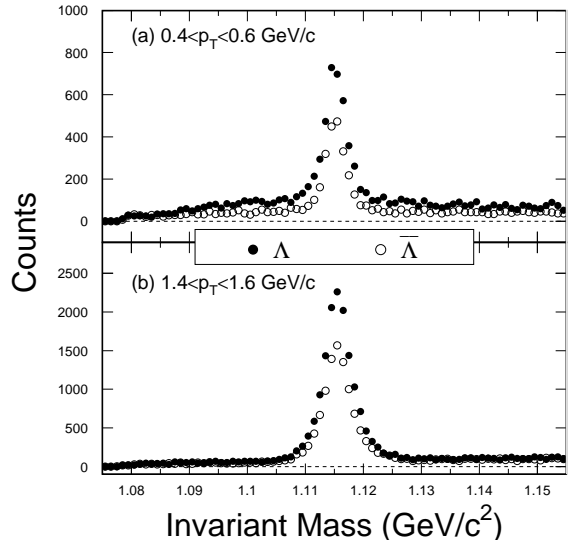,width=.49\textwidth}}
\vspace{-1.5cm}
\caption{Invariant mass distribution of $\Lambda$ (solid-circles) and
  $\overline{\Lambda}$ (open-circles) candidates from the data sample
analyzed including both central and minimum-bias
  collisions: (a) $0.4 < p_T < 0.6$ GeV/c; (b) $1.4 < p_T < 1.6$ GeV/c}
\label{fig:lmass}
\end{figure}

Figure~\ref{fig:lmass} shows the invariant mass distributions for
the reconstructed $\Lambda $ and $\overline{\Lambda}$ candidates in
$\left|y\right| <0.5$ for two typical $p_{T}$ bins from the data sample. 
The mass resolutions ($\sigma$) for reconstructed $\Lambda $ and
$\overline{\Lambda }$ are typically about $3$--$4$ MeV/c$^{2}$ based on
a Gaussian fit to the peak.  The background 
beneath the $\Lambda$($\overline{\Lambda}$) peak is dominated by
combinatoric pairs of charged particles. 
Decays of $K^{0}_{s}\rightarrow \pi ^{+}\pi ^{-}$ also
contribute to the smooth background due to pions misidentified as protons. 
The yield is obtained from the invariant mass
distribution in each transverse momentum ($p_{T}$) and rapidity ($y$)
bin, where the shape of the background near the $\Lambda
(\overline{\Lambda})$ peak is fit with a second order polynomial
function.  The signal to background ratios in the peak are greater than 6 for
all the measured $p_T$ bins.  Variations in the yield due to different
fits for the background have been included in the estimate of 
systematic errors.

The raw yield for each $p_{T}$-$y$ bin was then corrected for finite
detection efficiency. The overall correction factor was obtained from
a Monte Carlo embedding procedure.  $\Lambda $($\overline{\Lambda }$)
particles were generated using a STAR GEANT package for detector
simulation including all physical processes to take into account
effects such as scattering and absorption in the detector support
structures. A TPC Response Simulator (TRS) was used to generate TPC
information at the pad level for all charged tracks inside the TPC
from simulated $\Lambda$ and $\overline{\Lambda}$ decays. The TRS
includes all TPC resolution effects ranging from electron transport in the gas
to signal processing in the readout electronics.  The simulated data
were then embedded into real collision events from STAR detectors and
were analyzed with the standard STAR software package as real data.
The correction factor for each $p_{T}$-$y$ and multiplicity bin was
obtained from the number of $\Lambda $ ($\overline{\Lambda}$)
particles reconstructed from the embedded sample, divided by the
number generated. The combined acceptance and efficiency for  $\Lambda$ and
$\overline{\Lambda}$  ranges 
from $0.8\%$ to $5.8\%$ as a function of $p_{T}$ for the most central collision sample.

\begin{figure}[h]
\centering\mbox{
\psfig{figure=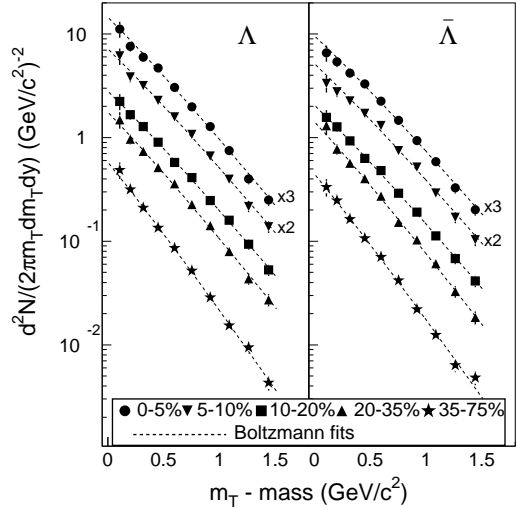,width=.42\textwidth}}
\vspace{0.4cm}
\caption{Transverse mass distributions of $\Lambda$ (left)
  and $\overline{\Lambda}$ (right) at mid-rapidity ($|y|$$<$$0.5$) 
  for selected centrality bins.  The
  dashed lines are Boltzmann fits.  Note that multiplicative factors have
  been applied to data from the two most central data sets for
  display.}
\label{fig:lbar_mt}
\end{figure}

\begin{figure}[h]
\centering\mbox{
\psfig{figure=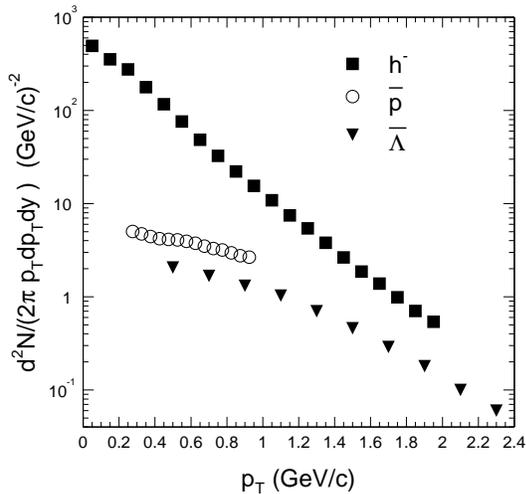,width=.4\textwidth}}
\caption{The mid-rapidity $\overline{\Lambda}$ ($|y|$$<$$0.5$) transverse 
momentum distribution from the top $5\%$ most central
  collisions. For comparison the distributions for negative hadrons
  ($d^2N/(2\pi p_{T})dp_{T}d\eta$, $|\eta|$$<$$0.1$) and anti-protons
  ($|y|$$<$$0.1$) for the similar centrality bin are included.
  Statistical errors are less than the size of the data points.}
\label{fig:pt_spec}
\end{figure}

\begin{table}
\begin{tabular}{|ll|c|c|c|c|c|} \hline
 \multicolumn{2}{|c|}{Centrality}  & 0--5$\%$ & 5--10$\%$ & 10--20$\%$ & 20--35$\%$ & 35--75$\%$ \\ \hline 
$dN/dy$ & $\Lambda $ & 17.0$\pm0.4 $ & 13.0$\pm0.3 $ & 10.1$\pm0.2 $ 
& 5.9$\pm0.2 $ & 1.61$\pm0.05 $ \\ \cline{3-7}
(Boltz) & $\overline{\Lambda }$ & 12.0$\pm0.3 $ & 9.6$\pm
0.3 $ & 7.4$\pm0.2 $ & 4.6$\pm0.1 $ & 1.26$\pm
0.04 $ \\ \hline 
$T_{B}$ & $\Lambda $ & 298$\pm5 $ & 304$\pm6 $ & 303$\pm6 $ 
& 289$\pm6 $ & 254$\pm5 $ \\ \cline{3-7} 
(MeV) & $\overline{\Lambda }$ & 312$\pm6 $ & 310$\pm6 $ & 
305$\pm6 $ & 280$\pm6 $ & 258$\pm5 $ \\ \hline\hline
$dN/dy$ &  $\Lambda $ & 17.4$\pm0.4 $ & 13.3$\pm0.3 $ 
& 10.4$\pm0.2 $ & 6.1$\pm0.2 $ & 1.66$\pm0.05 $ \\ \cline{3-7}
(exp fit) & $\overline{\Lambda }$ & 12.3$\pm0.3 $ & 9.8$\pm 0.3$ 
& 7.6$\pm0.2 $ & 4.7$\pm0.1$ & 1.30$\pm 0.04$ \\ \hline
$T_{E}$ & $\Lambda $ & 355$\pm8$ & 364$\pm9$ & 362$\pm8$ 
& 343$\pm8$ & 295$\pm7$ \\ \cline{3-7} 
(MeV)  & $\overline{\Lambda}$ & 374$\pm9$ & 373$\pm8$
& 366$\pm8$ & 331$\pm8$ & 301$\pm7$ \\ \hline
\end{tabular}
\caption{Fit parameters from Boltzmann and exponential fits of 
the $m_{T}$ spectra for $\Lambda$ and $\overline{\Lambda}$ at mid-rapidity ($|y|$$<$$0.5$). 
Only statistical errors are presented. The systematic errors on dN/dy and T are 
approximately $10\%$.}
\label{tab:dndy}
\end{table}

The measured $\Lambda $ spectra contain contributions from primordial
$\Lambda$, $\Sigma^{0}$ decays, and feed-down from
multiply-strange hyperons--notably $\Xi^{0}$ and $\Xi^{-}$. The
primordial $\Lambda $ and the $\Sigma^{0}$ decay products cannot be
separated in our analysis and have been treated as primary $\Lambda$.
We have investigated the feed-down from $\Xi $ hyperons by comparing
systematically the distribution of the DCA 
to the primary interaction vertex for the reconstructed
$\Lambda$ from real data versus simulated primary
$\Lambda$ and secondary $\Lambda$ from $\Xi$ decay. 
From this analysis, we estimate the
contributions of feed-down from multiply-strange hyperons, mostly
$\Xi$ and $\overline{\Xi}$ decays, to be approximately $(27\pm6)\%$ of the
measured $\Lambda$ and $\overline{\Lambda}$ yields respectively.
 
Figure~\ref{fig:lbar_mt} presents the $m_{T}$ spectra (invariant 
distributions) of $\Lambda $
and $\overline{\Lambda }$ for five selected centrality bins. 
Systematic errors from various methods of yield extraction,
reconstruction efficiencies and uncorrected sector-by-sector
variations in the TPC performance are included.  Both exponential
($e^{-(m_{T}-m)/T_{E}}$) and Boltzmann ($m_{T}e^{-(m_{T}-m)/T_{B}}$) 
functions, as well as a Gaussian ($e^{-p_{T}^{2}/(2\sigma^{2})}$) 
in $p_T$ have been used to fit the data, following the ansatz 
described in~\cite{starpbar}. 

All three functional forms were fit to the $\Lambda$ and $\overline{\Lambda}$ 
spectra used for this analysis. The slope parameter obtained from the exponential 
fit is systematically higher than that for the Boltzmann by approximately $40$--$65$ MeV. 
However, overall the integrated yields from all three fit functions are consistent 
within the statistical errors. The slope parameters and the $dN/dy$ for the Boltzmann 
and exponential fits are presented in Table~\ref{tab:dndy}. The Boltzmann form was 
adopted in Figure~\ref{fig:lbar_mt} because it typically provides a better $\chi^{2}$ 
and gives a reasonable description of the $m_{T}$ spectra over the entire range of
centrality and tranverse momenta investigated.

Within the systematic error, the slope parameters measured for the $\Lambda$ 
and $\overline{\Lambda}$ $m_{T}$ distributions are the same. There is a systematic 
increase in the slope parameters from approximately 254 MeV for the least 
central ($35$-$75\%$) to 312 MeV for the most central ($0$-$5\%$) bin. Similar behavior 
as a function of centrality is found for the $\overline{p}$ transverse mass 
distributions~\cite{starpbar}. Assuming the temperature at which particle interactions 
cease (the ``freezeout'' temperature~\cite{ssh}) is constant independent of 
collision centrality, the increase in the slope parameter may be interpreted as 
an increase in the collective radial velocity~\cite{na44,na49flow}.

A possible indication of hydrodynamic flow is the increase of the observed mean 
transverse momenta for various species with increasing particle mass. The transverse 
momentum distributions of negatively charged hadrons, as well as $\overline{p}$ 
and $\overline{\Lambda}$ are shown in Figure~\ref{fig:pt_spec}.
The $\overline{p}$ and $\overline{\Lambda}$ $p_{T}$ distributions are similar in shape 
in the region where they can be compared (below $1$ GeV/c), even though the data sets 
cover somewhat different ranges in $p_T$. Both distributions are qualitatively different from
and much less steep than the corresponding h$^{-}$ distribution, which is dominated by pions. 
Qualitatively similar behavior was observed in heavy ion collisions at the CERN Super Proton 
Synchrotron (SPS)~\cite{na44,na49flow,slope}. In general, the slope parameters for all species 
are observed to be larger at RHIC, which is attributed to an increase in the collective radial 
velocity as a function of center-of-mass energy~\cite{kaneta}.

Figure~\ref{fig:pt_spec} indicates that at higher $p_{T}$ ($p_T$$>$$1$~GeV/c) 
the ratio of $\overline{\Lambda}$ to negative hadrons increases rapidly. 
The baryon to meson ratio at RHIC for  $p_T$$>$$1$~GeV/c exceeds expectations from perturbative QCD 
inspired string fragmentation models which were tuned to fit $e^{+}e^{-}$ collision data and 
are the basis for modeling particle production
in hadronic collisions as well~\cite{jetset,gyulassy,hijingpr}.  
For example, the $\overline{\Lambda}$ to h$^{-}$ ratio is approximately 0.35 at $p_T$ of $2$ GeV/c.
Data from $e^{+}e^{-}$ collisions and calculations from string fragmentation models, however, 
indicate that although the baryon to meson ratio from
quark and gluon fragmentation increases as a function of Feynman $x$, 
the ratio never exceeds 0.2~\cite{hofmann}.
As mentioned above, a natural explanation for the increase at high $p_T$ would be a 
large collective radial flow at RHIC~\cite{ssh,slope}. Alternatively, it has also 
been suggested that the energy loss of high $p_T$ partons could modify 
the baryon to meson ratio~\cite{gyulassy}. To determine the exact dynamics 
that cause the relative enhancement of baryons to mesons at high
$p_T$, more experimental measurements over a larger $p_T$-range are needed.

\begin{figure}[h]
\centering\mbox{
\psfig{figure=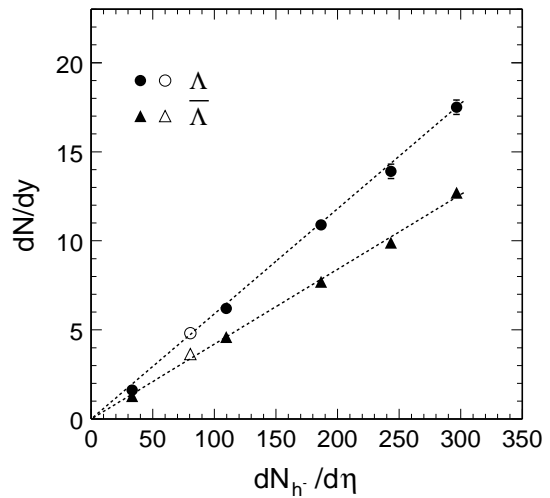,width=.4\textwidth}}
\caption{$\Lambda$ and $\overline{\Lambda}$ rapidity 
  density as a function of negative hadron multiplicity at
  mid-rapidity. The open symbols in the figure are data points from an
  independent analysis of minimum bias data set using event-mixing techniques.  Errors shown
  are statistical only.}
\label{fig:l_hm}
\end{figure}

The average ratio of $p_T$ integrated yield $\overline{\Lambda}$ to $\Lambda$ is $0.74\pm 0.04$ with no 
significant variation over the measured range of centrality. Given that there is a net 
excess of baryons at mid-rapidity, it is reasonable to conclude that there is more than 
one process contributing to $\Lambda$ production and that significant baryon number 
from the colliding beams is transported to $\Lambda$ hyperons at mid-rapidity. A question 
in that regard is why the shapes of $m_T$ spectra for $\Lambda$ and $\overline{\Lambda}$ are 
the same within errors. It has been suggested that significant rescattering of $\Lambda$ 
and $\overline{\Lambda}$ during the evolution of the collision can lead to 
equilibration~\cite{Rene,pbm01}. The nature of baryon number transport itself requires further study.

Figure~\ref{fig:l_hm} shows the $dN/dy$ of $\Lambda $ and
$\overline{\Lambda }$ from the Boltzmann fit as a function of 
the h$^{-}$ pseudo-rapidity
density \cite{starh}. 
At mid-rapidity the hyperon production is approximately proportional to
the primary h$^{-}$ multiplicity in Au+Au collisions at RHIC. The
dashed lines in the figure correspond to
$\Lambda=(0.054\pm0.001)$h$^{-}$ and
$\overline{\Lambda}=(0.040\pm0.001)$h$^{-}$ from a linear fit to the
data.  Similar centrality dependence of the lambda production was
observed at the SPS energies.  The $\overline{\Lambda}$
to h$^{-}$ ratio at RHIC is much larger than that at the SPS while the
$\Lambda$ to h$^{-}$ ratio is smaller at RHIC~\cite{WA97}. 
This may be understood from the fact that at the SPS, most of the observed $\Lambda$ 
hyperons carry baryon number transported from the colliding nuclei through associated 
production, rescattering and fragmentation processes. In this case, the yield of $\Lambda$ 
to h$^{-}$ ratio is larger than that at RHIC due to the relatively high fraction of 
the $\Lambda$ yield not resulting from pair production. Conversely, the increased 
importance of baryon pair production at RHIC energies must contribute to the observed increase 
of the $\overline{\Lambda}$ to h$^{-}$ ratio relative to the SPS measurement.

In conclusion, we have presented the first inclusive mid-rapidity
($|y|$$<$$0.5$) $\Lambda $ and $\overline{\Lambda }$ spectra as a
function of centrality from Au+Au collisions at the
$\sqrt{s_{_{NN}}}=130$ GeV energy. Salient features of the data include 1) large 
slope parameters of transverse mass spectra probably resulting from increased 
collective radial velocity at RHIC, 2) similar shapes for $\Lambda$ 
and $\overline{\Lambda}$ spectra despite the fact that a significant fraction of 
the $\Lambda$ hyperons at mid-rapidity carry baryon number from the incoming nuclei 
while the $\overline{\Lambda}$ hyperons are primarily pair produced, and 3) a significant 
increase in the $\overline{\Lambda}$ yield relative to negatively charged primary 
hadrons at moderate $p_T$ above $1$ GeV/c which cannot be described by existing perturbative 
QCD inspired string fragmentation models. The $p_T$ integrated rapidity densities 
of $\Lambda$ and $\overline{\Lambda}$ are approximately proportional to the 
number of negative hadrons at mid-rapidity.

{\bf Acknowledgments:} We wish to thank the RHIC Operations Group at
Brookhaven National Laboratory for their tremendous support and for
providing collisions for the experiment. This work was supported by
the Division of Nuclear Physics and the Division of High Energy
Physics of the Office of Science of the U.S. Department of Energy, the
United States National Science Foundation, the Bundesministerium fuer
Bildung und Forschung of Germany, the Institut National de la Physique
Nucleaire et de la Physique des Particules of France, the United
Kingdom Engineering and Physical Sciences Research Council, Fundacao
de Amparo a Pesquisa do Estado de Sao Paulo, Brazil, the Russian
Ministry of Science and Technology, the Ministry of Education of
China and the National Natural Science Foundation of China.




\begin{thebibliography}{99}
\bibitem{Blaizot} J.P. Blaizot, Nucl. Phys. {\bf A661}, 3c (1999) and
  references therein.
  
\bibitem{mult} B.B. Back {\it et al.}, Phys. Rev. Lett. {\bf 85}, 3100
  (2000).
  
\bibitem{phenix} K. Adcox {\it et al.}, Phys. Rev. Lett. {\bf 86}, 3500
  (2001).
  
\bibitem{starh} C. Adler {\it et al.}, Phys. Rev. Lett. {\bf 87}, 112303
  (2001).
  
\bibitem{ratio} C. Adler {\it et al.}, Phys. Rev. Lett. {\bf 86}, 4778
  (2001).
  
\bibitem{phobos} B.B. Back {\it et al.}, Phys. Rev. Lett., {\bf 87}, 102301 (2001).

\bibitem{brahms} I.G. Bearden {\it et al.}, Phys. Rev. Lett. {\bf 87}, 112305 (2001).
    
\bibitem{jetset} B. Andersson {\it et al.}, Physica Scripta {\bf 32},
  574 (1985).

\bibitem{Ellis} J. Ellis {\it et al.}, Phys. Lett. {\bf B233}, 223
  (1989).
  
\bibitem{Ledoux} W. Busza and R. Ledoux, Ann. Rev. of Nucl. and
  Part. Sci. {\bf 38}, 119 (1988).
  
\bibitem{Baryon} H.Z. Huang, Proceedings of Relativistic Heavy Ion
  Symposium APS Centennial Meeting '99, R. Seto, Editor. 3 (1999).

\bibitem{Khar} D. Kharzeev, Phys. Lett. {\bf B378}, 238 (1996).
  
\bibitem{Rafelski} J. Rafelski and B. M\"{u}ller, Phys. Rev. Lett. {\bf 48}, 1066 (1982); {\bf 56}, 2334(E) (1986).
  
\bibitem{E802} T. Abbott {\it et al.}, Phys. Rev. Lett. {\bf 64}, 847
  (1990).
  
\bibitem{Ahm} S. Ahmad {\it et al.}, Phys. Lett. B{\bf382}, 35
  (1996).
  
\bibitem{WA97} F. Antinori {\it et al.}, Eur. Phys. J. {\bf C11}, 79
  (1999).
  
\bibitem{NA49} H. Appelsh\"{a}user {\it et al.}, Phys. Lett. {\bf B444},
  523 (1998).

\bibitem{Koch} P. Koch {\it et al.}, Phys. Rep. {\bf 142}, 167 (1986).
  
\bibitem{Rene} R. Bellwied {\it et al.}, Phys. Rev. {\bf C62}, 054906
  (2000).
  
\bibitem{Sorge} H. Sorge, Phys. Rev. {\bf C52}, 3291 (1995).
  
\bibitem{Amelin} N. Amelin {\it et al.}, Phys. Lett. {\bf B306}, 312
  (1993).
  
\bibitem{Rapp} R. Rapp and E. Shuryak, Phys. Rev. Lett. {\bf 86}, 2980
  (2001).
  
\bibitem{Flow} K.H. Ackermann {\it et al.}, Phys. Rev. Lett. {\bf 86},
  402 (2001).


\bibitem{TPC} K.H. Ackermann {\it et al.}, Nucl. Phys. {\bf A661},
  681c (1999).
    
\bibitem{starpbar} C. Adler {\it et al.}, Phys.
  Rev. Lett. 87, 262302 (2001) .

\bibitem{ssh} E. Schnedermann, J. Sollfrank and U. Heinz, Phys. Rev.
  {\bf C48}, 2462 (1993) .

\bibitem{na44} I. Bearden {\it et al.}, Phys. Rev. Lett. {\bf 78}, 2080 (1997) .

\bibitem{na49flow} H. Appelsh\"{a}user {\it et al.}, Eur. Phys. J. {\bf C2}, 661 (1998) .

\bibitem{slope} H. van Hecke {\it et al.}, Phys. Rev. Lett. {\bf 81},
  5764 (1998).

\bibitem{kaneta} M. Kaneta and N. Xu, J. Phys. {\bf G27}, 589 (2001).

\bibitem{gyulassy} I. Vitev and M. Gyulassy, nucl-th/0104066.
    
\bibitem{hijingpr} X.N. Wang, Phys. Rep. {\bf 280}, 287 (1997) .

\bibitem{hofmann} W. Hofmann, Ann. Rev. Nucl. Part. Sci. {\bf 38}, 279 (1988).
  
\bibitem{pbm01} P. Braun-Munzinger, D. Magestro, K. Redlich, and J.
  Stachel, Phys. Lett. {\bf B518}, 41(2001).
    
\end{thebibliography}
\end{document}